Original Investigations

**Power-law distribution in Burst-suppression on electroencephalogram of dogs**

Chika Koyama, Taichi Haruna, Satoshi Hagihira, Kazuto Yamashita

Department of Small Animal Clinical Sciences, School of Veterinary Medicine, Rakuno Gakuen University, Hokkaido, 069-8591, Japan (C.K., K.Y.); Department of Information and Sciences, Tokyo Woman's Christian University, Tokyo, 167-8585, Japan (T.H.); Department of Anaesthesiology, Kansai Medical University, Osaka, 573-1010, Japan (S.H.)

*Correspondence to Chika Koyama, Department of Small Animal Clinical Sciences, School of Veterinary Medicine, Rakuno Gakuen University, 582 Bunkyodai- Midorimachi, Ebetsu, Hokkaido 069-8591, Japan
e-mail: s21641008@g.rakuno.ac.jp
Abbreviated Title: POWER-LAW IN BURST-SUPPRESSION OF DOG

**Abstract**

Burst-suppression (BS) is a reliable electroencephalogram (EEG) indicator of excessive deep anesthesia common in mammals. Automatic detection of BS with high accuracy will be important for anesthesia management. Since some intermittent events are known to follow a power-law, we investigated the power-law hypothesis in BS by comparing it with alternative functions focusing on flattish periods and developed a new method for detecting BS as an application of statistical model in dogs.

Young-adult 6 beagles and senior 6 beagles were anesthetized with sevoflurane 2.0%, 2.5%, 3.0%, 3.5%, 4.0%, and 5.0% and three of 64 sec EEG (256 Hz) from Fpz-T4 via scalp electrodes were recorded. Three thresholds for peak-to-peak voltage were set: mean value of peak-to-peak voltage at sevoflurane 2.0% in each dog ($AS_{2\%}$), 3μV, and 5μV. The subthreshold periods were discriminated as "$\tau$" events. We fitted the empirical probability distribution of $\tau$ by a power-law distribution and an exponential distribution. These two distributions were compared by the normalized log-likelihood ratio (**R**) test to see which distribution was better fit. **R**>0 with p<0.05 indicates that the power-law distribution is better fit than the exponential distribution in this study.

At sevoflurane 2.0%-3.0%, by any threshold, the exponential distribution became better fit in all dogs. The power-law distribution became better fit only when BS expressed on EEG. Especially by $AS_{2\%}$ threshold, **R**>0 with p<0.05 was in 11/12 dogs and |**R**|<2 with p>0.05 was in 1/12 dog at BS. No strict threshold was required for detection of onset of BS.

We showed a transition from exponential behavior to power-law behavior on the right tail of $\tau$ distributions in response to the appearance of suppression waves with increasing anesthetic. This will be a robust tool for BS detection that does not require a specific threshold in dogs.

## 1. Introduction

Burst-suppression (BS) pattern has been used as a reliable electroencephalogram (EEG) indicator of inactivated brain states for mammals observed at excessive deep anesthesia, coma, and hypothermia.[1-4] Excessive deep anesthesia is known to increase risk of postoperative delirium and cognitive impairment in human.[5-7] Therefore, automatic detection of BS with high accuracy would be important for anesthesia management. However, it is still insufficient.[8-10] This is especially noticeable in beagle dogs, where the suppression waves at onset of BS are often short (less than 0.5 sec) and hard to detect.[9] BS is defined as electrical pause ('suppression') alternating with high amplitude electrical activity ('Burst') and is detected as suppression ratio of a time domain EEG parameter.[11-13] The criteria for BS are specified so that the suppression ratio does not exceed zero at anesthesia concentrations below BS level diagnosed by visual inspection. Previously, we demonstrated that suppression ratio exceeded zero even at non-BS level if the criterion of its duration was shortened.[9] Indeed, looking at an enlarged view of EEG sampled at 256 Hz in dogs (figure 1E), it is confirmed that EEG contains a number of flattish periods where there is not much variance in voltage between successive peaks. We named that subthreshold period "$\tau$" (figure 1A).[14]

Power-law distributions: $p(k) = \propto k^{-\alpha}$ (α: constant), have been presented as a statistical model of Burst of BS.[15,16] Many other intermittent events in nature and in organisms, such as earthquake, snow avalanche, and neuronal avalanche, are known to follow a power-law,[17-19] characteristic for second-order phase transitions where the system is in a "critical" state.[20,21] Modeling distribution is extremely useful for understanding scientific activities by showing empirical rules and physical phenomena in a logical and objective way.[22] In brief, a power-law implies scale-free, in which rare events are observed that would not be allowed in a normal distribution, characterized by regularly varying with many events of small degree and a few events of large degree. This probability distribution and this cumulative probability distribution form a straight line on log-log scale. In addition, compared with an exponential distributions $q(k) = \propto \lambda^{-k}$ (λ: constant), the probability distribution shows more gradual decline in power-law.

On BS waveform in dogs, $\tau$ events have various scales from milliseconds to seconds (figure 1B, 1C), and hence deserve to investigate whether their statistical behavior has a scale-free-like characteristic. As far as we know, the phenomenon of suppression of BS has not been described by power-law yet. However, recent studies emphasize that tests of rigorous power-law hypotheses in empirical data are quite difficult.[23,24] Therefore, in this study, we first attempted to fit the probability distribution of $\tau$ to power law distribution and exponential distribution, following the method reported by Clauset et al..[25] We hypothesized that emergence of suppression waves would lead to dynamics close to power-law behavior in $\tau$ events. Next, we performed likelihood ratio tests of goodness-of-fit for these two distributions and examined whether an application of these statistic models could be a new method for detecting BS in dogs.

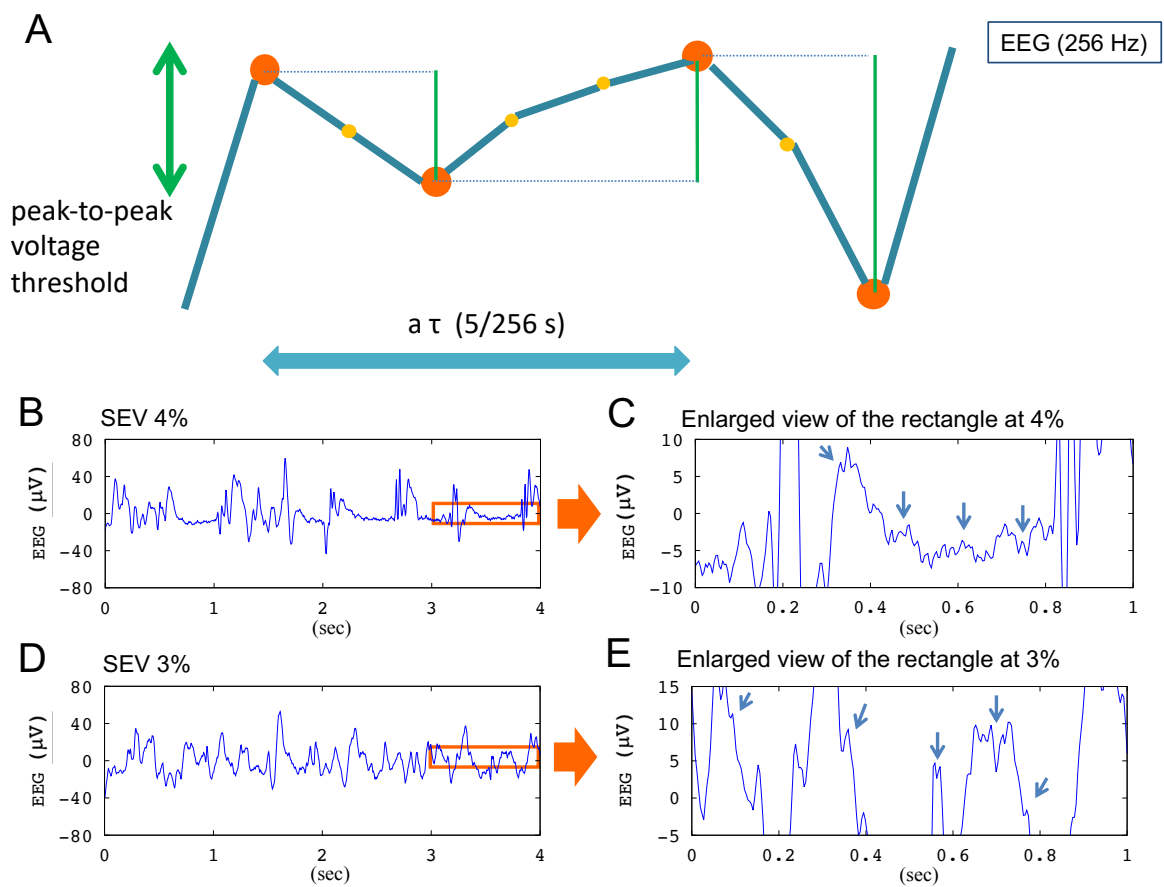

**Fig. 1.** Concept of τ (A): Illustration of a τ event using a schematic diagram of electroencephalographic (EEG) waveforms. A τ event is defined as a consecutive part on EEG in which the potential difference between adjacent peaks is within a given voltage threshold. Yellow and orange circles are EEG values sampled at 256Hz. The orange circle is the peak detected by the first derivative. (B): EEG waveform for 4 sec in dog during onset of Burst-suppression (BS) under 4% sevoflurane. (C): Enlarged view (1 sec, 20 μV) of the rectangle box at 4% sevoflurane. Fluctuations are observed on suppression waves (arrows). (D): EEG waveform for 4 sec in dog under 3% sevoflurane. (E): Enlarged view (1 sec, 20 μV) of the rectangle box at 3% sevoflurane. Minute waves (arrows) like a fragment of the suppression wave is observed here and there.

## 2. Materials and Methods

### 2.1. Animals

Experiments were conducted in 6 young-adult beagle dogs (2.5 ± 1.5 [mean ± SD] years old; 9.9 ± 0.9 kg) and 6 senior beagle dogs (10.1 ± 1.5 years old; 12.6 ± 1.4 kg) after approval by the Animal Care and Use Committee of Rakuno Gakuen University (Approval No. VH14B7). Each group comprised 3 males and 3 females. The dogs were cared for according to the principles of the "Guide for the Care and Use of laboratory Animals" prepared by Rakuno Gakuen University. All dogs were judged to be in good to excellent health based on physical examination, blood cell count and serum biochemical analysis. After the experiment, all dogs completed sevoflurane anesthesia and were awake. All experiments were performed non-invasively with no apparent inflammation and pain.

### 2.2. Anesthesia protocols

No prior medication was administered. General anesthesia was induced via mask using 8% sevoflurane (SevoFlo; DS Pharma Animal Health Co. Ltd., Osaka, Japan) in oxygen. After endotracheal intubation, all dogs were placed on mechanical ventilation (Nuffield Anesthesia Ventilation Series 200; Penlon, Abingdon Oxon, UK) with sevoflurane and oxygen in left lateral recumbency. Rocuronium bromide (ESLAX Intravenous; MSD Co., Inc., N.j., USA) was administered at a loading dose of 0.5 mg/kg and then infused constantly at a maintenance dose of 1 mg $\cdot$ kg$^{-1}$ $\cdot$ h$^{-1}$ with Ringer's lactate solution (5 mL $\cdot$ kg$^{-1}$ $\cdot$ h$^{-1}$) through the cephalic vein. Arterial blood pressure (ABP) was measured by connecting a catheter from the dorsal artery to a pressure transducer (BD DTXTM Plus DT-4812, Becton, Dickinson and Co., Fukushima, Japan) placed at the level of the mid-sternum. During anesthesia, the end-tidal partial pressure of carbon dioxide (ETCO$_2$) was maintained between 33 and 40 mmHg using intermittent positive pressure ventilation. Esophageal temperature was maintained at 37.5-38.0 °C using a warm air blanket. ABP, ETCO$_2$, esophageal temperature, heart rate (HR), and end-tidal concentration of sevoflurane (SEV) were monitored throughout the experiment (BP-608V, Omron Colin Co. Ltd., Tokyo, Japan). Just before each EEG recording, an arterial blood sample was collected and the partial pressure of arterial CO$_2$ was immediately analyzed to confirm a value between 33–40 mmHg. The anesthetic agent monitor was calibrated at the start of each experiment using a calibration kit.

### 2.3. Data acquisition and preprocessing

Anesthesia was maintained in order at 2.0%, 2.5%, 3.0%, 3.5%, 4.0% and 5.0% SEV in this study. The SEV 2.0% was determined based on the minimum alveolar concentration of SEV 1.3 ± 0.3 (%) that prevented voluntary response in 50% of dogs (MAC$_{awake}$).[26] Five min EEG sampled at 256 Hz were recorded at each SEV following a 20-min equilibration. The raw EEG data were from Fpz-T4 via scalp needle electrodes using an A-2000XP Bispectral index (BIS) monitor (ver. 3.21, Covidien-Medtronic, MN, USA). Simultaneously, raw signals were exported into EEG analysis software (Bispectral Analyzer for A2000).[27] Three EEG data packets of 64 sec were extracted for analysis from that 5 min data, avoiding the checking time of the BIS monitor. One-sec moving averages were subtracted from the data, which were divided into mutually non-overlapping 2 sec periods, and Welch's window function was applied. Subsequently, discrete Fourier transform was performed at 50 ± 1 Hz and 100 ± 1 Hz to remove noise caused by the alternating current power source.

The electrode junction was remodeled (Unique Medical Co. Ltd., Tokyo, Japan) to use needle electrodes (Disposable sub-dermal needle electrode NE-115B, Nihonkohden Co., Tokyo, Japan). After clipping the hair and disinfection with isopropyl alcohol solution, the needle electrodes were placed in a right front-temporal configuration, as previously described in dogs.[28] The electrode impedance was kept at 5 kΩ or less throughout the study. The notch filter was set at 50 Hz by BIS monitor. Muscle relaxants were administered in sufficient amounts for dogs to prevent myoelectric potential noise contamination of the EEG data.[29,30] All electrical appliances were powered through an uninterruptible power supply (APC Smart-UPS 1500, Schneider Electric, Paris, France) to reduce the noise of the AC power supply.

**2.4. $\tau$**

Three thresholds for peak-to-peak voltage were set: a mean value of peak-to-peak voltage at SEV 2.0% in each dog ($AS_{2\%}$), 3μV, and 5μV. After detecting peaks using the first derivative, continuous periods where peak-to-peak voltage did not exceed the threshold were discriminated as "$\tau$" periods (figure 1A). Using the total $\tau$s of the 64 sec three data at each concentration, the cumulative distribution function (CDF) of $\tau$ was computed for the statistical analysis.

**2.5. Judgment of the minimum SEV of BS occurrence by visual inspection ($SEV_{BS}$)**

Prior to the quantitative evaluation, for each dog, the minimum SEV at which BS was visually confirmed in all three of 64-sec EEG was determined as $SEV_{BS}$ (by C.K.). The BS criterion was set to one or more occurrences of flattish waves lasting more than 0.35 sec for 64 sec EEG waveform redrawn using gnuplot software.[9]

**2.6. Statistical analysis**

· Fitting power-law distribution and exponential distribution to CDF of $\tau$

The present study was conducted in the line with the statistical toolbox for detecting power-law behavior introduced by Clauset et al..[25] First, the data of CDF were fitted by a power-law distribution. The probability density function of power-law model was defined as follows;

$$p(\tau) = \frac{\alpha - 1}{\tau_{min}} \left(\frac{\tau}{\tau_{min}}\right)^{-\alpha},$$

where α>1 and $\tau \geq \tau min$. The scaling parameter α was estimated by the method of maximum likelihood. The lower bound on power-law behavior $\tau_{min}$ was determined by the method using the Kolmogorov-Smirnov (KS) statistic. The likelihood function is defined as follows;

$$L_1 = \prod_{i=1}^{m} p(\tau_i),$$

where the m is the number of data $\tau_i$ satisfying $\tau_i \geq \tau_{min}$. The KS statistic is defined as follows;

$$KS = \max_{\tau \geq \tau min} |S(\tau) - p(\tau, \alpha)|,$$

where S($\tau$) is the empirical CDF for $\tau$ in the range above $\tau_{min}$. For each $\tau \geq \tau min$, α was repeatedly calculated by maximum likelihood estimate. The α and $\tau_{min}$ was computed so that the KS statistic was the smallest. The $\tau_{min}$ was set so that more than half of number of $\tau$ were analyzed for fit. Next, the data satisfying $\tau \geq \tau_{min}$ were fitted by an exponential distribution. The probability density function of exponential model is defined as follows:

$$q(\tau) = \lambda \exp(\lambda(\tau_{min} - \tau)),$$

where $\tau \geq \tau min$. The parameter λ was estimated by the method of maximum likelihood as follows;

$$L_2 = \prod_{i=1}^{m} q(\tau_i),$$

where the m is the number of data $\tau_i$ satisfying $\tau_i \geq \tau_{min}$.

· Comparison of goodness-of-fit using the logarithm of likelihood ratio

Basically, the one with the higher likelihood is better fit. The two distributions were compared using the normalized log- likelihood ratio test as follows;

$$\mathbf{R} = \log \frac{L1}{L2},$$

which is positive or negative depending on which distribution is better fit, or zero if the two are equal. To quantitatively evaluate that the value of $\mathbf{R}$ is significantly away from zero, the probability (p-value) that a given the value of $\mathbf{R}$ was measured when the true value of $\mathbf{R}$ is close to 0, was estimated using a method proposed by Voung [31]. The p-value was given by the standard deviation σ of $\mathbf{R}$ as follows;

$$\mathbf{R} = \sum_{i=1}^{m} (\ln p(\tau_i) - \ln q(\tau_i)) = \sum_{i=1}^{m} (l_{1i} - l_{2i})$$

where $l_{1i} = p(\tau_i)$ or $l_{2i} = \ln q(\tau_i)$ is as the log-likelihood for a single measurement $\tau i$ within each distribution.

$$\sigma^2 = \frac{1}{m} \sum_{i=1}^{m} [(l_{1i} - l_{2i}) - (\overline{l_1} - \overline{l_2})]^2$$

with

$$\overline{l_1} = \frac{1}{m} \sum_{i=1}^{m} l_{1i}, \qquad \overline{l_2} = \frac{1}{m} \sum_{i=1}^{m} l_{2i}$$

Using σ of $\mathbf{R}$, p-value denotes as follows;

$$p = \frac{1}{\sqrt{2\pi n \sigma^2}} \left[ \int_{-\infty}^{-|R|} e^{\frac{-t^2}{2n\sigma^2}} dt + \int_{-|R|}^{\infty} e^{\frac{-t^2}{2n\sigma^2}} dt \right]$$

$$= erfc\left(\frac{|R|}{\sqrt{2n}\sigma}\right)$$

and,

$$erfc(Z) = 1 - erf(Z)$$

$$= \frac{2}{\sqrt{\pi}}\int_{Z}^{\infty} e^{-t^2}$$

is the complementary function of Gaussian error.

The p-value less than 0.05 was considered significant. Data were analyzed by a program written in C. The graphical outputs were generated using gnuplot (Version 5.1, http://www.gnuplot.info/).

## 3. Results

### 3.1. SEV$_{BS}$ and AS$_{2\%}$ threshold

Results of SEV$_{BS}$ and threshold values of AS$_{2\%}$ are summarized in Table 1. SEV$_{BS}$ was 4.0%-5.0% in young-adult beagles, and 3.5%-4.0% in senior beagles. In all dogs, SEV 5% inhalation achieved BS levels on EEG; In No. 1-3 of young-adult dogs and all seniors, EEG was dominated only by BS pattern. Especially in No. 11 and 12, EEG was almost a flat wave with 5~6 or 1~3 Burst waves in 64 sec, respectively. In No. 4-6 dogs, EEG showed fast-and-slow waves with BS waves. Mean ± SD of AS$_{2\%}$ threshold was 6.7 ±1.5 (μV) in young-adult beagles and 5.4 ± 1.7 (μV) in senior beagles.

Table 1. The minimum end-tidal sevoflurane concentration of Burst-suppression expression (SEVBS) and a mean value of peak-to peak voltage at 2.0 % sevoflurane (AS$_{2\%}$) in young-adult (YA) and senior (S) dogs.

| group | Dogs | SEV$_{BS}$ (%) | AS$_{2\%}$ (μV) |
|---|---|---|---|
| YA | No.1 | 5.0 | 5.8 |
| | No.2 | 5.0 | 6.4 |
| | No.3 | 4.0 | 9.5 |
| | No.4 | 5.0 | 4.9 |
| | No.5 | 5.0 | 6.0 |
| | No.6 | 5.0 | 7.3 |
| S | No.7 | 3.5 | 6.0 |
| | No.8 | 3.5 | 8.7 |
| | No.9 | 4.0 | 3.7 |
| | No.10 | 4.0 | 4.1 |
| | No.11 | 3.5 | 5.5 |
| | No.12 | 4.0 | 4.2 |

### 3.2. A typical example of results in No.8 by AS$_{2\%}$ threshold of 8.7 μV

By visual inspection of EEG waveforms in No.8, SEV$_{BS}$ was defined at SEV 3.5% (figure 2A). As in the other dogs, frequency and duration of suppression wave emergence tended to increase at BS level with increasing SEV. At SEV 5.0%, EEG was dominated by a pattern of spike-like Burst wave and suppression wave of about 1 sec. Visually on a double logarithmic graph, CDF of $\tau$ was nearly identical between SEV 2.0% and 3.0% of non-BS level (figure 2B). Conversely, at onset of BS level, the right tail curve of CDF was shallower than that at non-BS level, with the right edge of CDF extended to over $\tau$ = 0.30 sec. With increasing SEV, the descent curve became shallower and sharp decline in the right end became more pronounced around $\tau$ = 1 sec. The power-law and exponential distributions derived as a statistical model based on CDF are shown by green and blue auxiliary lines, respectively. It was observed that CDF shifted from exponential behavior to power-law behavior around SEV$_{BS}$ with SEV increasing, as all other dogs (Supplementary figure 1). The results of fitting the two distribution are summarized in figure 2C. The value of **R** was negative with p<0.001 between SEV 2.0-3.0 %, and was positive with p<0.05 between SEV 3.5-5.0 %. The value of **R** decreased at SEV 5.0% with almost regular BS pattern, compared to that at 4.0% with irregular BS pattern mixed with slow waves. The parameter λ was almost constant between SEV 2.0-3.0 % and decreased in a concentration-dependent manner between SEV 3.5-5.0 %. The parameter α was 2.3-2.5 between SEV 2.0-3.5 %, 1.9 at 4.0 %, and 1.3 at 5.0 %.

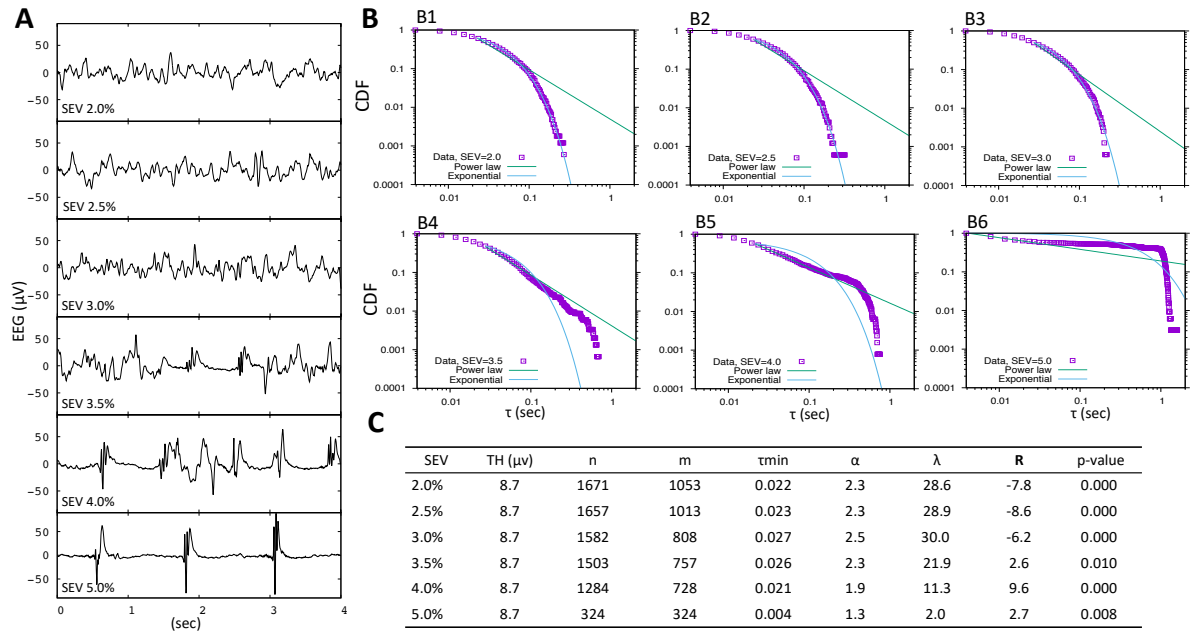

| SEV | TH (μv) | n | m | τmin | α | λ | R | p-value |
|---|---|---|---|---|---|---|---|---|
| 2.0% | 8.7 | 1671 | 1053 | 0.022 | 2.3 | 28.6 | -7.8 | 0.000 |
| 2.5% | 8.7 | 1657 | 1013 | 0.023 | 2.3 | 28.9 | -8.6 | 0.000 |
| 3.0% | 8.7 | 1582 | 808 | 0.027 | 2.5 | 30.0 | -6.2 | 0.000 |
| 3.5% | 8.7 | 1503 | 757 | 0.026 | 2.3 | 21.9 | 2.6 | 0.010 |
| 4.0% | 8.7 | 1284 | 728 | 0.021 | 1.9 | 11.3 | 9.6 | 0.000 |
| 5.0% | 8.7 | 324 | 324 | 0.004 | 1.3 | 2.0 | 2.7 | 0.008 |

**Fig.2.** One example of the results in a 12.1-year-old male beagle (No.8). (A) The electroencephalographic (EEG) waveform of 4 sec duration during 2.0%, 2.5%, 3.0%, 3.5%, 4.0%, and 5.0% of sevoflurane (SEV) anesthesia. The onset of Burst-suppression (BS) was at 3.5% SEV. The suppression waves occasionally appeared where the fast and slow waves were predominant at 3.5% and 4.0%. The frequency and duration of the suppression waves were prolonged by increasing SEV. At 5.0% SEV, EEG showed only Burst-suppression waves on EEG. As reference, the suppression ratio was 3% at 3.5% SEV, 20% at 4.0% SEV, and 77% at 5.0% SEV by manual inspection. (B) Cumulative distribution function (CDF) of τ (purple squares) and its fittings by a power-law distribution (green line) and an exponential distribution (blue line). The τ was discriminated as a subthreshold period on EEG measured in dogs under 2.0% (B1), 2.5% (B2), 3.0% (B3), 3.5% (B4), 4.0% (B5), and 5.0% (B6) SEV anesthesia. The threshold (TH) was set as the average of peak-to-peak differences at SEV 2.0% (8.7μV). First, the data were fitted by a power-law distribution $p(\tau) = \alpha \cdot 1/\tau_{min} (\tau/\tau_{min})^{-\alpha}$. The scaling parameter α was estimated by the method of maximum likelihood. The lower bound on power-law behavior τmin was determined by the method using the Kolmogorov-Smirnov statistic. Next, the data satisfying τ ≥ τmin were fitted by an exponential distribution $q(\tau) = \lambda \exp(\lambda(\tau_{min} - \tau))$. The parameter λ was estimated by the method of maximum likelihood. (C) Fitting parameters of CDF of τ and the results of the normalized log-likelihood ratio test. **R**> 0 with p<0.05 indicates that the power-law distribution is better fit than the exponential distribution, where **R** is the normalized log-likelihood ratio calculated from the data satisfying τ ≥ τmin between the power-law distribution p(τ) and the exponential distribution q(τ). Conversely, **R**< 0 with p<0.05 indicates that the exponential distribution is better fit than the power-law distribution. n is the number of τ. m is the number of data satisfying τ ≥ τmin.

### 3.3. The results of R

The results of **R** are summarized in Table 2. The color of the number identifies which of the power-law distribution (red; **R**>2) or the exponential distribution (black; **R**<-2) fits significantly better. When |**R**| was less than 2 (blue), there was no significant difference in goodness-of-fit between those two distributions in this study. The results of fitting the two distribution to the data of CDF, the number of data and the scaling parameters are summarized in Table 3 for each group.

Regardless of the threshold, **R** indicated that the power-law distribution was significantly better fit only at BS level of over SEV 3.5%. In addition, **R** did not indicate that the exponential distribution was significantly better fit at BS level, except by 3μV threshold in No.11 and No.12. In contrast, between SEV 2.0% and 3.0%, by any threshold, **R** indicated that the exponential distribution was significantly better fit, and the pattern of CDF was nearly identical with λ above 20 in all dogs.

There were two patterns in which |**R**| <2 was less than 2, as follow; First, at non-BS level in No.1 and No.2, the visual inspection of EEG waveforms revealed the appearance of suppression waves lasting 0.35 sec in one or two of three data. In these cases, the right end of CDF slightly elevated and extended. The λ was almost unchanged, compared to the other CDF at non-BS level. Second, in No.11 and No.12, EEG waveforms showed almost flat waves. In these cases, the pattern of CDF had a large curve with λ decreasing between 2 and 0.

**R** was negative at BS level in No.11 and No.12, where 3μV threshold was adapted for almost flat-wave EEG data. In these cases, the λ was between 2 and 8.

Table 2. The results of **R** that is the normalized log-likelihood ratio calculated from the data satisfying $\tau \geq \tau_{min}$ between the power-law distribution $p(\tau)$ and the exponential distribution $q(\tau)$ in young adult (No.1 - No.6) and senior (No.7 - No.12) dogs anesthetized with 2.0%-5.0% sevoflurane (SEV). The mean value of peak-to-peak voltage at SEV 2.0% for each individual (AS2%), 3μV, and 5μV were set as the threshold (TH). **R** >2 shown in red indicates that the power-law distribution is significantly better fit than the exponential distribution. **R** <-2 shown in black indicates that the exponential distribution is significantly better fit than the power-law distribution. |**R**|<2 shown in light blue indicates no significant difference. The SEV at which Burst-suppression level was confirmed is blue in the square. Statistical significance was ascribed to a p value less than 0.05. The right blue rectangles indicate the appearance of Burst-suppression on the EEG.

| SEV (%) | TH | No.1 | No.2 | No.3 | No.4 | No.5 | No.6 | No.7 | No.8 | No.9 | No.10 | No.11 | No.12 |
|---|---|---|---|---|---|---|---|---|---|---|---|---|---|
| 2.0% | AS2% | -8.8 | -8.6 | -6.5 | -8.0 | -9.2 | -8.6 | -6.6 | -7.8 | -9.2 | -6.2 | -10.2 | -8.4 |
|  | 3 μV | -6.8 | -4.6 | -5.7 | -6.0 | -5.7 | -6.7 | -9.2 | -7.6 | -7.1 | -8.8 | -8.4 | -9.2 |
|  | 5 μV | -7.3 | -9.9 | -6.1 | -7.6 | -9.1 | -8.0 | -7.1 | -7.1 | -9.4 | -8.5 | -7.1 | -8.3 |
| 2.5% | AS2% | -7.5 | -9.1 | -9.2 | -6.2 | -9.0 | -8.6 | -8.7 | -8.6 | -7.7 | -9.3 | -5.9 | -7.1 |
|  | 3 μV | -8.8 | -4.2 | -5.7 | -5.4 | -8.1 | -6.9 | -6.5 | -3.6 | -7.6 | -8.9 | -5.0 | -6.1 |
|  | 5 μV | -7.6 | -8.2 | -7.7 | -6.7 | -7.1 | -8.7 | -9.1 | -6.9 | -10.2 | -7.8 | -6.6 | -6.7 |
| 3.0% | AS2% | -6.5 | -2.5 | -6.8 | -4.9 | -6.5 | -3.4 | -6.0 | -6.2 | -8.2 | -6.3 | -5.5 | -7.4 |
|  | 3 μV | -7.7 | -3.1 | -6.6 | -6.5 | -4.1 | -4.9 | -4.1 | -4.5 | -7.5 | -6.6 | -5.4 | -6.2 |
|  | 5 μV | -7.2 | -2.3 | -5.8 | -5.7 | -6.4 | -4.0 | -6.0 | -4.6 | -10.2 | -8.7 | -7.0 | -7.6 |
| 3.5% | AS2% | -1.1 | -5.7 | -2.7 | -5.6 | -4.5 | -6.0 | 4.4 | 2.6 | -2.2 | -3.4 | 6.6 | -2.5 |
|  | 3 μV | -0.7 | -4.0 | -4.0 | -4.8 | -6.5 | -7.4 | 5.4 | 5.3 | -2.2 | -1.2 | 10.5 | -2.2 |
|  | 5 μV | 0.2 | -4.4 | -1.2 | -4.9 | -5.7 | -5.4 | 5.1 | 4.6 | -5.2 | -4.0 | 7.2 | -2.0 |
| 4.0% | AS2% | -1.7 | -0.1 | 15.6 | -5.9 | -7.0 | -6.8 | 4.6 | 9.6 | 8.2 | 14.4 | 11.7 | 4.5 |
|  | 3 μV | -1.3 | -4.0 | 4.3 | -4.9 | -2.5 | -7.5 | 7.4 | 8.7 | 8.8 | 14.7 | -6.2 | 5.3 |
|  | 5 μV | -1.1 | -1.4 | 13.8 | -5.7 | -5.9 | -8.1 | 6.0 | 10.5 | 5.8 | 12.7 | 10.6 | 3.3 |
| 5.0% | AS2% | 4.1 | 5.8 | 5.9 | 10.7 | -0.6 | 5.3 | 5.3 | 2.7 | 3.2 | 13.1 | 2.4 | -1.3 |
|  | 3 μV | -0.1 | -1.5 | 7.1 | 7.2 | 1.1 | 3.7 | 7.6 | 12.3 | 0.5 | 11.0 | -14.2 | -5.0 |
|  | 5 μV | 1.3 | 4.0 | 9.3 | 10.0 | 0.2 | 5.5 | 4.1 | 14.1 | 7.8 | 11.3 | 0.7 | -0.8 |

Table 3. The number of τ in 192 s EEG (Nτ), the percentage of τ satisfying τ ≥ τmin (Nτ'/Nτ), the lower bound on the behaviors (τmin), the scaling parameters for power-law (α) and exponential (λ) function, and **R** of the result of the log-likelihood ratio test in young-adult (YA) and senior (S) dogs anesthetized with sevoflurane (SEV) multiples by the peak-to-peak voltage thresholds (TH) of a mean value at SEV 2.0% in each dogs (AS2%), 3μV, and 5μV. The results were shown as mean ± SD.

| TH | SEV (%) | Nτ | | Nτ'/Nτ (%) | | τmin (ms) | | α | | λ | | R | |
|---|---|---|---|---|---|---|---|---|---|---|---|---|---|
| | | YA | S | YA | S | YA | S | YA | S | YA | S | YA | S |
| AS2% | 2.0 | 2451 ± 294 | 2113 ± 220 | 58 ± 5 | 57 ± 5 | 20 ± 2 | 22 ± 1 | 2.5 ± 0.1 | 2.4 ± 0.1 | 40 ± 3 | 35 ± 3 | -8.3 ± 0.9 | -8.1 ± 1.4 |
| | 2.5 | 2276 ± 182 | 2076 ± 248 | 56 ± 4 | 55 ± 5 | 20 ± 2 | 23 ± 2 | 2.5 ± 0.1 | 2.4 ± 0.1 | 41 ± 5 | 33 ± 3 | -8.3 ± 1.1 | -7.9 ± 1.1 |
| | 3.0 | 2067 ± 127 | 1951 ± 193 | 52 ± 2 | 53 ± 2 | 21 ± 1 | 23 ± 2 | 2.6 ± 0.1 | 2.5 ± 0.1 | 44 ± 4 | 35 ± 3 | -5.1 ± 1.7 | -6.6 ± 0.9 |
| | 3.5 | 1958 ± 140 | 1754 ± 288 | 55 ± 6 | 52 ± 2 | 19 ± 2 | 21 ± 3 | 2.6 ± 0.1 | 2.3 ± 0.2 | 45 ± 2 | 27 ± 9 | -4.3 ± 1.8 | 0.9 ± 3.8 |
| | 4.0 | 1756 ± 211 | 1317 ± 468 | 54 ± 6 | 67 ± 16 | 18 ± 2 | 15 ± 6 | 2.5 ± 0.3 | 1.9 ± 0.3 | 42 ± 14 | 13 ± 6 | -1.0 ± 7.9 | 8.8 ± 3.6 |
| | 5.0 | 1694 ± 297 | 415 ± 286 | 67 ± 16 | 92 ± 18 | 13 ± 5 | 181 ± 395 | 2.2 ± 0.5 | 1.5 ± 0.3 | 35 ± 27 | 3 ± 2 | 5.2 ± 3.3 | 4.2 ± 4.4 |
| 3μV | 2.0 | 2812 ± 383 | 2438 ± 184 | 51 ± 1 | 56 ± 5 | 12 ± 1 | 15 ± 3 | 2.9 ± 0.1 | 2.6 ± 0.0 | 98 ± 13 | 64 ± 14 | -5.9 ± 0.8 | -8.4 ± 0.8 |
| | 2.5 | 2537 ± 216 | 2390 ± 239 | 53 ± 2 | 52 ± 2 | 12 ± 1 | 16 ± 2 | 2.9 ± 0.1 | 2.6 ± 0.2 | 95 ± 9 | 61 ± 15 | -6.5 ± 1.6 | -6.3 ± 1.7 |
| | 3.0 | 2202 ± 158 | 2147 ± 215 | 53 ± 5 | 52 ± 3 | 12 ± 1 | 16 ± 3 | 2.8 ± 0.1 | 2.7 ± 0.1 | 93 ± 11 | 62 ± 15 | -5.5 ± 1.6 | -5.7 ± 1.2 |
| | 3.5 | 2013 ± 198 | 1927 ± 273 | 57 ± 7 | 52 ± 2 | 11 ± 2 | 16 ± 2 | 2.8 ± 0.1 | 2.4 ± 0.2 | 94 ± 12 | 40 ± 11 | -4.6 ± 2.1 | 2.6 ± 4.8 |
| | 4.0 | 1895 ± 217 | 1635 ± 332 | 53 ± 4 | 67 ± 17 | 12 ± 1 | 12 ± 4 | 2.7 ± 0.3 | 1.9 ± 0.3 | 82 ± 26 | 21 ± 8 | -2.7 ± 3.6 | 6.4 ± 6.3 |
| | 5.0 | 2132 ± 264 | 931 ± 451 | 56 ± 10 | 83 ± 20 | 12 ± 2 | 66 ± 105 | 2.7 ± 0.6 | 1.7 ± 0.3 | 77 ± 40 | 6 ± 4 | 2.9 ± 3.4 | 3.4 ± 9.7 |
| 5μV | 2.0 | 2637 ± 187 | 2067 ± 107 | 53 ± 5 | 51 ± 1 | 17 ± 2 | 25 ± 5 | 2.6 ± 0.1 | 2.5 ± 0.1 | 57 ± 10 | 36 ± 10 | -8.0 ± 1.2 | -7.9 ± 0.9 |
| | 2.5 | 2408 ± 115 | 2013 ± 81 | 56 ± 6 | 56 ± 5 | 16 ± 2 | 23 ± 6 | 2.6 ± 0.1 | 2.4 ± 0.1 | 56 ± 7 | 34 ± 10 | -7.7 ± 0.7 | -7.9 ± 1.3 |
| | 3.0 | 2146 ± 100 | 1907 ± 90 | 52 ± 2 | 54 ± 4 | 17 ± 2 | 22 ± 4 | 2.7 ± 0.1 | 2.5 ± 0.1 | 58 ± 9 | 37 ± 11 | -5.2 ± 1.6 | -7.4 ± 1.8 |
| | 3.5 | 1999 ± 137 | 1701 ± 186 | 51 ± 1 | 54 ± 4 | 17 ± 2 | 21 ± 4 | 2.7 ± 0.1 | 2.3 ± 0.2 | 60 ± 9 | 25 ± 6 | -3.5 ± 2.2 | 0.9 ± 4.8 |
| | 4.0 | 1796 ± 149 | 1304 ± 398 | 57 ± 8 | 67 ± 17 | 15 ± 3 | 16 ± 7 | 2.5 ± 0.3 | 1.9 ± 0.3 | 52 ± 17 | 13 ± 5 | -1.4 ± 7.2 | 8.1 ± 3.3 |
| | 5.0 | 1876 ± 263 | 452 ± 309 | 64 ± 11 | 93 ± 14 | 13 ± 2 | 98 ± 209 | 2.3 ± 0.4 | 1.4 ± 0.1 | 46 ± 30 | 3 ± 2 | 5.1 ± 3.7 | 6.2 ± 5.4 |

## 4. Discussion

This study was started from the observation of EEG waveforms at BS level as follows; Subthreshold events of "$\tau$" with various scales from milliseconds to seconds are alternating with above-threshold events. Next, focusing on the micro-amplitude $\tau$ as scale-free intermittent events, we investigated power-law hypothesis for $\tau$ on BIS using statistical models in dogs under sevoflurane anesthesia. The optimal algorithm for estimating the parameters was referred to the report by Closet et al..[25] This study presented **R**s, the results of log-likelihood ratio tests between power-law distribution and exponential distribution. As hypothesized, the right tail curve of CDF was shallow and its end extended depending on generation of suppression wave on BS. Consequently, **R** fluctuated from a negative value to a positive value with emergence of suppression wave as SEV increased. This indicates that CDF of $\tau$ shows a transition from exponential behavior to power-law behavior with BS appearance. At the same time, we demonstrated that this statistical model of $\tau$ could be used to determine the deep level of anesthesia in which BS appeared on the EEG in dogs. Above all, in any dog, no strict threshold setting was required for discrimination of non-BS or BS by **R**.

### 4.1. Comparison with existing method of BS detection

Although $\tau$ and suppression wave are common in waves below the voltage threshold, $\tau$ is all subthreshold waves regardless of length of duration, whereas suppression wave is a wave with more than the minimum visible length. Therefore, the number of detections was quite different between those two waves. For BS detection, our proposed method utilized the distribution of $\tau$ rather than the total percentage of the suppression wave as implemented in clinically practiced monitor.[8,12] Previous reports on machine-generated suppression ratio pointed out the low accuracy of detecting BS.[8,9] If the existing methods implement only one type of criteria (e.g., waves lasting more than 0.5 sec below 5 μV), it will be hard to detect BS accurately with individual differences. In a previous study using the same EEG data of senior dogs in this study, we also reported poor results with less than 20% detection of suppression ratio by BIS monitor at both SEV 3.5% and 4.0%.[9] Conversely, even if the suppression ratio is greater than 0, BIS value does not always show deep anesthesia level.[11,32] Our proposed method in this study, even though by various thresholds, could detect near the onset of BS with excellent sensitivity and specificity. This will be due to the fact that the pattern of distribution of $\tau$ does not change much at a certain range of thresholds.

### 4.2. Differences by age

In general, anesthetic requirement and amplitude of EEG decrease with age.[33-35] The differences in the results of the present study between young-adult and senior groups would be consistent with these general theories. First, SEV$_{BS}$, anesthetic concentration required to reach BS as an endpoint of anesthetic action, tended to be lower in the older group (young-adult dogs vs senior dogs = 4.0%-5.0% vs 3.5%-4.0%). Next, the $\tau$min prolonged not only by increasing the value of threshold but also with aging. For example, comparing the mean $\tau$min of young-adult and senior groups at non-BS level, it was 12 msec vs 16 msec by 3 µV threshold, and 17 msec vs 23 msec by 5 µV threshold, respectively (table 3). The higher the threshold, the more likely $\tau$ will be detected continuously. The $\tau$min was prolonged in older group. This would be due to the effect of decreasing amplitude with age. The decrease in amplitude with aging could be physiologically explained by an increase in resistance due to tissue fibrosis. Finally, the mean λ tended to be inversely proportional to the value of threshold and age. This could be explained by steeper curve of CDF for population with shorter duration of $\tau$. The present method showed a trend by age, however, there was not much variation among each dog. On the other hand, in four dogs of the young-adult group (except No. 4 and No. 6), the fluctuation of suppression waves of BS was clearly greater than that in senior dogs (Supplementary figure 2). Because of this phenomenon, criteria of BS that we previously determined using senior beagles[9] were not suitable in young-adult beagles. In this study, regardless of the variation in amplitude with age and in the settings of three threshold values, detection of BS by **R** could be generally successful.

**4.3. Power-law distribution in $\tau$ of BS in dogs**

With increasing SEV, **R** changed from negative (at non-BS) to positive (at BS) through zero (at pre or onset of BS), and finally close to zero (at flat waves). The results of **R** reflected variations of CDF of $\tau$ with suppression wave appearance. During non-BS level, CDF was nearly identical with a steep curve with λ above 20 and **R** indicated that exponential distribution was better fit. Appearance of suppression waves would lead to detection of long-lasting $\tau$s and change in the curve of CDF becoming shallower. Accordingly, **R** showed a better fit from exponential distribution to power-law distribution, then closer to somewhere in between. Only by 3 µV threshold, **R** was negative at BS with almost flat EEG waves in No.11 and No.12. This could be due to the fact that the right tail curve of CDF became steeper by shorter $\tau$ collective with that small threshold. Nevertheless, the λ value was contained in an increase up to 8, which was not as high as in non-BS. In those cases, both **R** and λ can be used to determine whether or not the BS level. When we tested at 1µV threshold, we found the similar trend of negative **R** at BS also in the other dogs (Data not shown). Thus, in order to discriminate BS using **R**, a threshold greater than 3µV would be suitable in beagle dogs.

Although this study did not directly examine whether CDF of $\tau$ follows a power-law, the value of parameter $\alpha$ in this study and the physiology of the suppression waves generation are consistent with the features of power-law distribution. Previous reports have shown that the empirical data were most likely to follow power-law distributions when the scaling parameter $\alpha$ was between 2 and 3,[23,36] which is consistent with the values of $\alpha$ in this study. The generation of BS, the endpoint of anesthetics effect on EEG, could be considered a critical situation: With increasing anesthetic concentration, once BS appeared on EEG, the suppression wave tended to appear more frequently and last longer, as previously reported.[3,11] The anesthetics acts on γ-aminobutyric acid A receptors to open the chloride channel and hyperpolarize.[37] Subsequently, the excessive hyperpolarization leads to Burst generation by endogenous pacemakers in some thalamic neurons through blocking input from thalamus to cortical neuron.[38,39] This is like a response to a critical situation in the biological system and is consistent with the reported characteristics of the power-law.[21] These results suggest that not only the reported Burst[15,16] of BS but also $\tau$ may follow the power-law.

**4.4. Limitations**

One of the most important limitations is that this study of mathematical models was conducted on a small number of dogs. Further studies should verify that power-law distribution of CDF and accuracy of BS discrimination by **R** in many other dogs and species. The next important limitation is the sampling frequency of EEG. The current analysis focused on $\tau$ with msec unit of time could not be performed on data greater than 256Hz recorded. For reference, the minimum $\tau$ used for the fitting was around 20 msec, which included about eight sampled data. Considering that frequency analysis requires data collection at twice the maximum frequency to be analyzed, that does not seem so bad. In addition, we also confirmed that the current analysis was also valid for data down-sampled to 128Hz: The number of $\tau$ detected decreased to about 80% and the pattern of change in CDF of $\tau$ for SEV was almost the same. However, further study is required to examine the dynamics of $\tau$ with respect to the sampling frequency.

**4.5. Conclusions**

We examined CDF of $\tau$ at with or without the appearance of BS on EEG in dogs anesthetized with sevoflurane, and estimated a statistical model of the exponential or power-law distributions to represent $\tau$ phenomena using the log-likelihood ratio (**R**) test. This study showed that the right tails of CDF shifted from an exponential behavior to a power-law behavior in response to emergences of BS. In addition, these statistical models did not require exact thresholds for each dog. Hence, **R** value could be a robust tool to predict BS on EEG in dogs.

**Acknowledgements:** The authors would like to acknowledge Jun Tamura, DVM, PhD (Department of Veterinary Medicine, Rakuno Gakuen University, Ebetsu, Japan) and the students who supported this experiment at Rakuno Gakuen University.

**Funding Statement**: Support was provided solely from institutional sources.

**Conflicts of Interest**: Competing Interest: The authors declare no competing interests

**Author's contribution:** C.K.: This author helped design study, proposal of new parameter, collect and analyze data, and write the manuscript; T.H.: This author helped create analysis program, proposal of statistic models, analyze data, and revise the manuscript; S.H.: This author helped design study, proposal of BSA for BIS software, and revise the manuscript. K.Y.: This author helped design experimental study, collect data, and revise the manuscript.

References

1. Brown EN, Lydic R, Schiff ND: General anesthesia, sleep, and coma. N Engl J Med 2010; 363: 2638-50

2. Steriade M, Amzica F, Contreras D: Cortical and thalamic cellular correlates of electroencephalographic burst-suppression. Electroencephalogr Clin Neurophysiol 1994; 90: 1-16

3. Swank RL, Watson CW: Effects of barbiturates and ether on spontaneous electrical activity of dog brain. J Neurophysiol 1949; 12: 137-60

4. Michenfelder JD, Milde JH: The relationship among canine brain temperature, metabolism, and function during hypothermia. Anesthesiology 1991; 75: 130-6

5. Soehle M, Dittmann A, Ellerkmann RK, Baumgarten G, Putensen C, Guenther U: Intraoperative burst suppression is associated with postoperative delirium following cardiac surgery: a prospective, observational study. BMC Anesthesiol 2015; 15: 61

6. Fritz BA, Kalarickal PL, Maybrier HR, Muench MR, Dearth D, Chen Y, Escallier KE, Ben Abdallah A, Lin N, Avidan MS: Intraoperative Electroencephalogram Suppression Predicts Postoperative Delirium. Anesth Analg 2016; 122: 234-42

7. Watson PL, Shintani AK, Tyson R, Pandharipande PP, Pun BT, Ely EW: Presence of electroencephalogram burst suppression in sedated, critically ill patients is associated with increased mortality. Crit Care Med 2008; 36: 3171-7

8. Muhlhofer WG, Zak R, Kamal T, Rizvi B, Sands LP, Yuan M, Zhang X, Leung JM: Burst-suppression ratio underestimates absolute duration of electroencephalogram suppression compared with visual analysis of intraoperative electroencephalogram. Br J Anaesth 2017; 118: 755-761

9. Koyama C, Haruna T, Hagihira S, Yamashita K: New criteria of burst suppression on electroencephalogram in dogs anesthetized with sevoflurane. Res Vet Sci 2019; 123: 171-177

10. Hagihira S, Okitsu K, Kawaguchi M: Unusually low bispectral index values during emergence from anesthesia. Anesth Analg 2004; 98: 1036-8, table of contents

11. Bruhn J, Bouillon TW, Shafer SL: Bispectral index (BIS) and burst suppression: revealing a part of the BIS algorithm. J Clin Monit Comput 2000; 16: 593-6

12. Rampil IJ: A primer for EEG signal processing in anesthesia. Anesthesiology 1998; 89: 980-1002

13. Drover D, Ortega HR: Patient state index. Best Pract Res Clin Anaesthesiol 2006; 20: 121-8

14. Koyama C, Haruna T, Hagihira S, Yamashita K: New EEG parameters correlated with sevoflurane concentration in dogs: tau and burst. 2019; arXiv:1910.02768

Search

15. Roberts JA, Iyer KK, Finnigan S, Vanhatalo S, Breakspear M: Scale-free bursting in human cortex following hypoxia at birth. J Neurosci 2014; 34: 6557-72

16. Rae-Grant AD, Kim YW: Type III intermittency: a nonlinear dynamic model of EEG burst suppression. Electroencephalogr Clin Neurophysiol 1994; 90: 17-23

17. Beggs JM, Plenz D: Neuronal avalanches in neocortical circuits. J Neurosci 2003; 23: 11167-77

18. Birkeland K, Landry C: Power-laws and snow avalanches. Geophysical Research Letters - GEOPHYS RES LETT 2002; 29

19. Gutenberg B, Richter CF: Frequency of earthquakes in California*. Bulletin of the Seismological Society of America 1944; 34: 185-188

20. Beggs JM, Timme N: Being critical of criticality in the brain. Front Physiol 2012; 3: 163

21. Bak P, Tang C, Wiesenfeld K: Self-organized criticality: An explanation of the 1/f noise. Phys Rev Lett 1987; 59: 381-384

22. George E.P. Box NRD: Empirical Model-Building and Response Surfaces (Wiley Series in Probability and Statistics). Wiley 1987; 1st Edition

23. Broido AD, Clauset A: Scale-free networks are rare. Nat Commun 2019; 10: 1017

24. Voitalov I, van der Hoorn P, van der Hofstad R, Krioukov D: Scale-free networks well done. Physical Review Research 2019; 1: 033034

25. Clauset A, Shalizi CR, Newman MEJ: Power-Law Distributions in Empirical Data. Siam Review 2009; 51: 661-703

26. Hofmeister EH, Brainard BM, Sams LM, Allman DA, Cruse AM: Evaluation of induction characteristics and hypnotic potency of isoflurane and sevoflurane in healthy dogs. Am J Vet Res 2008; 69: 451-6

27. Hagihira S, Takashina M, Mori T, Mashimo T, Yoshiya I: Practical issues in bispectral analysis of electroencephalographic signals. Anesth Analg 2001; 93: 966-70, table of contents

28. Campagnol D, Teixeira Neto FJ, Monteiro ER, Beier SL, Aguiar AJ: Use of bispectral index to monitor depth of anesthesia in isoflurane-anesthetized dogs. Am J Vet Res 2007; 68: 1300-7

29. Sakata H, Ishikawa Y, Ishihara G, Oyama N, Itami T, Umar MA, Sano T, Yamashita K: Effect of sevoflurane anesthesia on neuromuscular blockade produced by rocuronium infusion in dogs. J Vet Med Sci 2019; 81: 425-433

30. Dugdale AH, Adams WA, Jones RS: The clinical use of the neuromuscular blocking agent rocuronium in dogs. Vet Anaesth Analg 2002; 29: 49-53

31. Vuong QH: Likelihood Ratio Tests for Model Selection and Non-Nested Hypotheses. Econometrica 1989; 57: 307-333

32.	Morimoto Y, Monden Y, Ohtake K, Sakabe T, Hagihira S: The detection of cerebral hypoperfusion with bispectral index monitoring during general anesthesia. Anesth Analg 2005; 100: 158-61

33.	Yamashita K, Iwasaki Y, Umar MA, Itami T: Effect of age on minimum alveolar concentration (MAC) of sevoflurane in dogs. J Vet Med Sci 2009; 71: 1509-12

34.	Magnusson KR, Scanga C, Wagner AE, Dunlop C: Changes in anesthetic sensitivity and glutamate receptors in the aging canine brain. J Gerontol A Biol Sci Med Sci 2000; 55: B448-54

35.	Schultz B, Schultz A, Grouven U, Zander I, Pichlmayr I: [Changes with age in EEG during anesthesia]. Anaesthesist 1995; 44: 467-72

36.	Barabasi AL, Albert R: Emergence of scaling in random networks. Science 1999; 286: 509-12

37.	Steinbach JH, Akk G: Modulation of GABA(A) receptor channel gating by pentobarbital. J Physiol 2001; 537: 715-33

38.	Schaul N: The fundamental neural mechanisms of electroencephalography. Electroencephalogr Clin Neurophysiol 1998; 106: 101-7

39.	Schaul N, Green L, Peyster R, Gotman J: Structural determinants of electroencephalographic findings in acute hemispheric lesions. Ann Neurol 1986; 20: 703-11